\documentstyle[12pt]{article}
\input epsf.sty
 1

\catcode`\@=11 
\makeatletter
\def\@seccntformat#1{\csname the#1\endcsname.\hskip 1em}

\makeatother
\begin{document}

\thispagestyle{empty}
\begin{flushright}

{\footnotesize\renewcommand{\baselinestretch}{.75}
           SLAC-PUB-8259\\
          September 1999\\
}
\end{flushright}


\begin{center}
{\large \bf  Recent Tests of QCD at SLD$^{*}$}

\vspace {1.0cm}

 {\bf David Muller}

\vspace {0.3cm}

 {\bf Representing The SLD Collaboration$^{**}$}

\vspace {0.1cm}

Stanford Linear Accelerator Center \\
Stanford University, Stanford, CA~94309 \\

\vspace{1.0cm}
{\bf Abstract}
\end{center}

\renewcommand{\baselinestretch}{1.2}

We present selected results on strong interaction physics from the SLD
experiment at the SLAC Linear Collider.
We report on several new studies of 3- and 4-jet 
hadronic $Z^0$ decays, in which jets are identified as quark, antiquark or
gluon.
The 3-jet $Z^0\!\rightarrow \! b\bar{b}g$ rate is 
sensitive to the $b$-quark mass; prospects for measuring $m_b$
are discussed.
The gluon energy spectrum is measured over the full kinematic range,
providing an improved test of QCD and limits on anomalous $bbg$ couplings.
The parity violation in $Z^0\!\rightarrow \! b\bar{b}g$ decays is
consistent with electroweak theory plus QCD.
New tests of T- and CP-conservation at the $bbg$ vertex are performed.
A new measurement of the rate of gluon splitting into $b\bar{b}$ pairs yields
$g_{b\bar{b}}\!=\!0.0031\!\pm\!0.0007 (stat.)\! \pm\! 0.0006 (syst.)$ (Preliminary).
We also present a number of new results on jet fragmentation into identified
hadrons.
The $B$ hadron energy spectrum is measured over the full
kinematic range using a new, inclusive technique,
allowing stringent tests of predictions for its shape 
and a precise measurement of 
$\left< x_B \right> \!=\!0.714\!\pm\!0.005 (stat.)\!\pm\!0.007 (syst.)$ (Preliminary).
A detailed study of correlations in rapidity $y$ between pairs of identified
$\pi^\pm$, $K^\pm$ and p/$\bar{\rm p}$ confirms that strangeness and baryon
number are
conserved locally, and shows local charge conservation between meson-baryon
and strange-nonstrange pairs.
Flavor-dependent long-range correlations are observed for all
combinations of these hadron species, yielding new
information on leading particle production.
The first study of correlations using rapidities signed such that $y>0$
corresponds to the quark direction provides additional new insights into
fragmentation, including the first direct observation of baryon number ordering
along the $q\bar{q}$ axis.

\vspace{0.5cm}
\begin{center}
{\it Presented at the 1999 International Euroconference on Quantum
Chromodynamics,\\
7--13 July, 1999, Montpellier, France.}
\end{center}

\vfil

\noindent
$^*$This work was supported in part by DOE grant DE-AC03-76SF00515.

\eject

\section{Introduction}
We present an overview of a number of recent measurements from the SLD
experiment, using hadronic decays of $Z^0$ bosons produced in $e^+e^-$
annihilations.
First we present a number of precision tests of QCD in the perturbative regime
using 3- and 4-jet final states.
The goal here is sensitivity to radiative corrections to the reaction
$e^+e^- \!\!\rightarrow \! Z^0/\gamma \!\!\rightarrow \! q\bar{q}(g)$ induced by
known effects,
such as the large mass of the $b$-quark, or by new physics.
Such effects are expected at the few percent level, so the effects of
higher order gluon radiation must be understood or suppressed at this level.
Since we do not observe partons directly, rather the jets of particles
into which they fragment, it is essential to understand the properties of these
jets.
Here we present two new, detailed studies of jet formation involving 
identified hadrons measured over a wide energy range.

A key to improving upon existing studies of multijet events has been the
identification of the parton species that initiated a particular jet, or 
the identification of the primary $q$ flavor in
$e^+e^- \!\rightarrow \! q\bar{q}$.
For example, the strong coupling $\alpha_s$ can be measured from $R_3$, the
fraction of 3-jet $e^+e^- \!\!\rightarrow \! q\bar{q}g$ events, but the
precision is limited by theoretical uncertainties to the level of $\sim$5\%.
However one can test the fundamental anzatz of flavor independence of $\alpha_s$
by comparing $R_3$ in events of different flavors so that
many uncertainties cancel.
The ratio $r^b = R_3^b / R_3^{uds}$ has been measured with sufficient
precision that it is sensitive to the mass of the $b$-quark, $m_b$, 
introducing additional theoretical and experimental uncertainties.
Here we review our latest measurements \cite{asflv} of $r^b$, and discuss
prospects for the precise measurement of $m_b$.

Experimental studies of the structure of 3-jet events have typically used 
energy and angle distributions of energy-ordered jets.
Since the gluon is expected to be the lowest-energy jet in most events,
this suffices to confirm the $q\bar{q}g$ origin of
such events and to determine the gluon spin \cite{gspin}.
The identification of the three jets in such events would allow
more complete and stringent tests of QCD.
Here we present a study \cite{confbbgx} of 3-jet final states in which two of
the jets are tagged as $b$ or $\bar{b}$ jets.
The remaining jet is tagged as the gluon jet and its energy spectrum studied
over the full kinematic range.
Adding a tag of the charge of the $b$ or $\bar{b}$ jet,
and exploiting the high electron beam polarization of the SLC,
we measure \cite{confbbgs} two angular asymmetries sensitive to parity
violation in the $Z^0$ decay, and also construct new tests of T- and
CP-conservation at the $bbg$ vertex.

The rate of secondary heavy flavor production via gluon splitting,
$g \!\rightarrow \! c\bar{c}$, $g \!\rightarrow \! b\bar{b}$ is a
sensitive test of QCD, as it is suppressed strongly by the mass of the heavy
quark, but is still expected to be the dominant source of secondary heavy
hadrons.
Here we present a measurement of the $g\!\rightarrow \! b\bar{b}$ rate \cite{confgbb}
that is complementary to other measurements in this rapidly emerging field.

The study of events containing $b$/$\bar{b}$ quarks is especially useful,
both as important input into measurements such as electroweak parameters
($R_b$ and $A_b$)
in $Z^0$ decays and bottom production in hadron-hadron collisions,
and also as a sensitive probe of any new physics that
couples more strongly to heavier quarks.

The process of jet formation is not understood quantitatively, due
to the difficulty of perturbative calculations in this soft regime.
A number of phenomenological models have been developed, and it is
essential to understand the properties of jets empirically
in order to test these models and encourage theoretical development.
Since jets are used in many precision tests of electroweak and strong
physics (e.g. those described below), and will constitute both the largest
signal for and the background to any heavy particles discovered in the
future, our understanding, even if only through models, must be as complete as
possible.

The production of heavy hadrons from heavy primary quarks is relatively easy to
calculate perturbatively due to the cutoff introduced by the large quark mass.
A number of calculations and model predictions for the energy spectrum of
bottom hadrons now exist and await precise testing.
Experimental studies of the $B$-hadron spectrum have been limited by the
efficiency for reconstructing the energies $E_B$ of individual $B$
hadrons with good resolution, especially for low-energy $B$ hadrons.
Here we present a study \cite{confxb} of the $E_B$ distribution
using a novel kinematic technique and only charged tracks.
The high efficiency and good resolution for all $E_B$, results in a
measurement covering the full kinematic range.

Lighter identified particles are also an active field of study.
The production of strange particles and baryons is of particular interest as
they must be produced in strange-antistrange or baryon-antibaryon pairs, and the
mechanism of their pair production can yield insights into the
fragmentation process.
Previous studies of rapidity correlations have shown that the conservation of
such quantum numbers is predominantly local or short-range,
i.e. the two particles are produced close together in the jet phase space.
There is also evidence for long-range correlations from the leading
particles in the two hemispheres of an event that contain the primary quark
and antiquark.

Here we present detailed studies \cite{confcorl} of short- and
long-range correlations between identified $\pi^\pm$, $K^\pm$ and
p/$\bar{\rm p}$ that improve substantially upon previous measurements.
In addition, we use the SLC beam polarization to tag the quark hemisphere in
each event and study for the first time rapidities signed such that positive
(negative) rapidity corresponds to the (anti)quark hemisphere.
Ordered differences in signed rapidity provide further unique insights into jet
fragmentation, in particular being sensitive to local ordering of
quantum numbers along the $q\bar{q}$ axis.

Unless otherwise noted, these measurements use the entire SLD hadronic
sample of 550,000 events recorded from 1993 to 1998.
Each exploits some of the unique feaures of the SLC/SLD program, such as the
highly longitudinally polarized electron beam, small transverse beam size, and 
excellent vertex detection and particle identification.
The magnitude of the beam polarization averaged 73\%, providing high
sensitivity to parity violating observables and a quark hemisphere tag of
73\% purity.
The collision region typically measured 0.8 $\mu$m vertically by 1.8 $\mu$m
horizontally, which with the vertex detector gives a transverse
impact parameter resolution of
$\sigma_{\delta} = 8\oplus 29/p\sin^{3/2}\theta$ $\mu$m, providing excellent
heavy- and light-flavor tagging.
We use a topological algorithm \cite{topo} that finds
secondary vertices cleanly and efficiently, and also measures the flight
direction independent of the momenta of the tracks in the vertex.
Cutting on the vertex mass provides $B$ (charmed) hadron samples of up to 98\%
(70\%) purity.
In addition we consider the number of tracks in an event or jet,
$n_{sig}^{(m)}$, that have a normalized transverse impact parameter wrt
the interaction point (IP) of $\delta/\sigma_\delta > m$.
Heavy hadron decays tend to give $n_{sig}^{(m)} > 0$, and
the absence of such tracks provides an efficient light-flavor tag.

\section{Flavor Independence and $m_b$}

\begin{figure}
\vspace {-0.1cm}
   \epsfxsize=2.9in
   \begin{center}\mbox{\epsffile{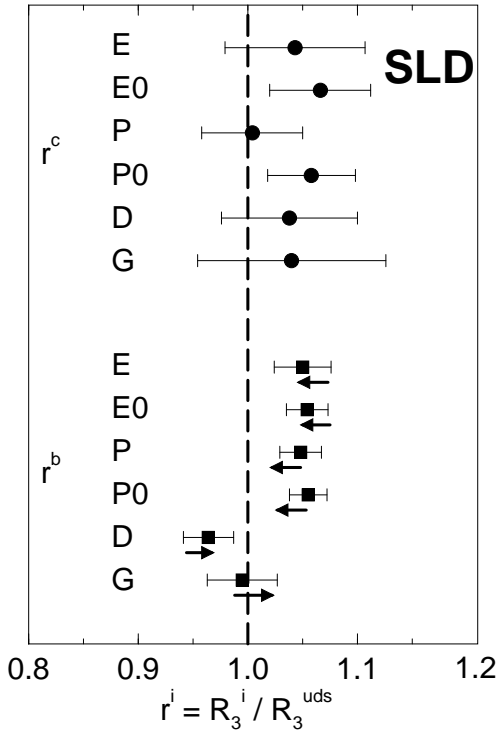}}\end{center}
\vspace {0.6cm}
\caption{ \baselineskip=12pt \label{asmb}
The measured 3-jet rate ratios $r^c$ and $r^b$ for six different jet finding
algorithms.
The predictions of the NLO QCD calculation of [10] are
shown as the arrows whose tails (heads) represent an $m_b(m_Z)$ value of
3.5 (2.5) GeV/c$^2$.
  }
\end{figure}

Well contained hadronic events \cite{evsel} are selected and jets are counted in
each using six collinear- and infrared-safe algorithms,
with a set of optimized $y_{cut}$ values.
A combination~\cite{asflv} of the event $n^{(2)}_{sig}$ and the
masses and momenta of the secondary vertices (if any) in each jet
is used to tag each event as light($uds$)-, $c$-, or $b$-flavor.
The 3-jet rates $R_3$ in these three subsamples are unfolded for the effects of
hadronization, detector efficiency and tagging efficiencies and biases, and the
ratios $r^c=R_3^c/R_3^{uds}$ and $r^b=R_3^b/R_3^{uds}$ are shown in
fig.~\ref{asmb}.

All six $r^c$ values are consistent with unity, however only one $r^b$ value is
within one standard deviation of unity.
Furthermore the deviation from unity
is different for different algorithms, and the six results show more scatter
than would be expected given the high statistical correlation between them.
This can be understood as an effect of the $b$-quark mass, $m_b$, which
leads to a suppression of collinear gluon radiation, and also affects
the jetfinders directly, as they use the mass in different ways
in their clustering metrics.
There are several recent NLO calculations \cite{rbbbu,rbothr} of the expected
shift in $r^b$, which give consistent predictions.  Those of \cite{rbbbu} are
shown as the arrows on fig.~\ref{asmb} for
values of the running mass of $m_b(m_Z)=3.0\pm 0.5$ GeV/c$^2$;
qualitative agreement with the data is evident.

We therefore used this prediction to extract ratios of the strong coupling:
\begin{eqnarray*}
\alpha_s^c / \alpha_s^{uds} & = & 1.036 \pm 0.043 (stat.) ^{+0.041}_{-0.045} (syst.) ^{+0.020}_{-0.018} (theory) \\
\alpha_s^b / \alpha_s^{uds} & = & 1.004 \pm 0.018 (stat.) ^{+0.026}_{-0.031} (syst.) ^{+0.018}_{-0.029} (theory),
\end{eqnarray*}
using one-sixth of the data sample.
The experimental uncertainty is expected to improve substantially when the
entire sample is analyzed, and a dominant theoretical error on the latter
ratio is from the uncertainty on $m_b(m_Z)$.

Recently, a number of groups \cite{mbbbmou,mbother} have tried an alternative
strategy of using these or similar data to measure
$m_b(m_Z)$ assuming flavor independence of $\alpha_s$.
The experimental precision is already very good,
however it is important to consider possible theoretical uncertainties.
For example Brandenburg et al. have shown \cite{mbbbmou}, using an updated
calculation, that our six values of $r^b$ are inconsistent if only experimental
errors are considered.
Postulating an additional 2\% theoretical uncertainty due to uncalculated higher
order terms, they extract
\[
m_b(m_Z) = 2.56 \pm 0.27 (stat.) ^{+0.28}_{-0.38} (syst.) ^{+0.49}_{-1.48} (theory) \;\; {\rm GeV/c}^2,
\]
\noindent
where the latter error has large contributions from uncertainties in the
hadronization correction as well as from the postulated theoretical uncertainty.
If our theoretical knowledge is improved in these two areas, then this promises
to provide a precise measurement of the $b$-quark mass.

\section{The Gluon Energy Spectrum}

Hadronic events with exactly 3 jets (JADE algorithm, $y_{cut}=0.02$)
are selected.
Jet energies $E_i$ are calculated from the interjet angles \cite{evsel}, and
the jets are energy ordered: $E_1\! >\! E_2\! >\! E_3$.
We require $n^{(3)}_{sig} \geq 2$ in exactly two of the three jets, and the
remaining jet is tagged as the gluon jet.
This yields 8196 events with an estimated purity of correctly tagged gluon
jets of 91\%.
In 3.0\% (12.9\%) of these events, jet 1(2), the (second) highest energy jet,
is tagged as the gluon jet, giving coverage over the full kinematic range.

\begin{figure}[t]
   \epsfxsize=4.0in
   \begin{center}\mbox{\epsffile{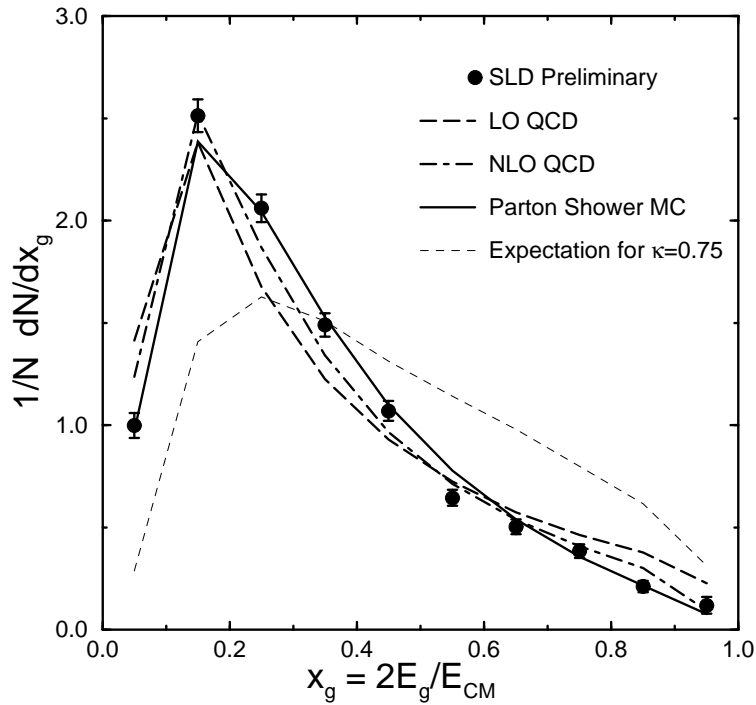}}\end{center}
\vspace {1.1cm}
\caption{ \baselineskip=12pt \label{xg}
The fully corrected scaled gluon energy distribution (dots).
The predictions of leading- and next-to-leading-order QCD
and of a parton shower simulation are shown as the dashed, dot-dashed and
solid lines, respectively.
The thin dashed line shows the prediction for an anomalous chromomagnetic
coupling at the bbg vertex with relative strength 0.75.
  }
\end{figure}

The distribution of the scaled gluon energy $x_g=2E_g/\sqrt{s}$ is
corrected \cite{confbbgx} for non-$b\bar{b}g$ and mistag backgrounds,
selection efficiency and resolution.
The fully corrected spectrum, fig.~\ref{xg}, shows the
expected falling behaviour with increasing $x_g$.
It is cut off at low $x_g$ by the finite $y_{cut}$ value.
Also shown are the predictions of first and second order QCD \cite{rbbbu};
both describe the data in general, but not in detail.
The prediction of the JETSET \cite{jetset} parton shower simulation is also
shown and reproduces the data.
We thus confirm the prediction of QCD,
although higher order effects are clearly important in the intermediate gluon
energy range, $0.2<x_g<0.4$.

The $x_g$ spectrum is particularly sensitive to the presence of an
anomalous chromomagnetic term in the strong interaction Lagrangian.
A fit of the theoretical prediction \cite{rizzo} including an anomalous
term parametrized by a relative coupling $\kappa$, yields a value of
$\kappa = -0.01 \pm 0.05$ (Preliminary), consistent with zero, and
corresponding to 95\% C.L. limits on such contributions to the $bbg$ coupling
of $-0.11<\kappa <0.08$.

\section{Parity Violation in 3-Jet $Z^0$ Decays}

We now consider two angles, the polar angle of the quark with respect to
the electron beam direction $\theta_q$, and
the angle between the quark-gluon and quark-electron beam planes
$\chi\!=\!\cos^{-1} (\hat{p}_q \!\times\! \hat{p}_g) \!\cdot\! (\hat{p}_q \!\times\! \hat{p}_e)$.
The cosine $x$ of each of these angles should be distributed as
$1\!+\!x^2\!+\!2 A_P A_Z x$, where the $Z^0$ polarization $A_Z\!=\!(P_e\!-\!A_e)/(1\!-\!P_eA_e)$
depends on that of the $e^-$ beam $P_e$, and $A_e$ and $A_P\!=\!A_0A_q$ are
predicted by QCD plus electroweak theory.

Three-jet events (Durham algorithm, $y_{cut}=0.005$) are selected and energy
ordered.
The 14,658 events containing a secondary vertex with mass above 1.5 GeV/c$^2$ in
any jet are kept, having an estimated $b\bar{b}g$ purity of 85\%.
We calculate the momentum-weighted charge of each jet $j$,
$Q_j=\Sigma_i q_i |\vec{p}_i \cdot \hat{p}_j |^{0.5}$, using the charge $q_i$
and momentum $\vec{p}_i$ of each track $i$ in the jet.
We assume that the highest-energy jet is not the gluon, and
tag it as the $b$ ($\bar{b}$) if $Q = Q_1 - Q_2 - Q_3$ is negative
(positive).
We define the $b$-quark polar angle by
$\cos\theta_b = -{\rm sign}(Q)(\hat{p}_e \cdot \hat{p}_1)$.

\begin{figure}
   \epsfxsize=3.1in
   \begin{center}\mbox{\epsffile{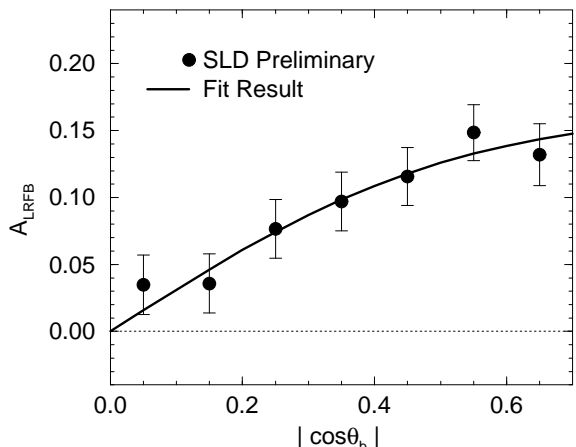}}\end{center}
\vspace {0.6cm}
\caption{ \baselineskip=12pt \label{pviol}
Left-right-forward-backward asymmetry of the $b$-quark
polar angle in 3-jet $Z^0$ decays.
The line is the result of a fit.}
\end{figure}

The left-right-forward-backward asymmetry $A_{LRFB}^b$ in
$\cos\theta_b$ \cite{confbbgs} is shown as a function of $|\cos\theta_b|$ in
fig.~\ref{pviol}.
The clear asymmetry increases with increasing $|\cos\theta_b|$ in the
expected way.
A fit to the data yields an asymmetry parameter of
$A_P\! = \!0.91\! \pm \!0.05 (stat.)\! \pm 0.06 (syst.)$
(Preliminary), consistent with the QCD prediction of
$A_P = 0.93 A_b = 0.87$.

\begin{figure}
\vspace {-0.3cm}
   \epsfxsize=3.1in
   \begin{center}\mbox{\epsffile{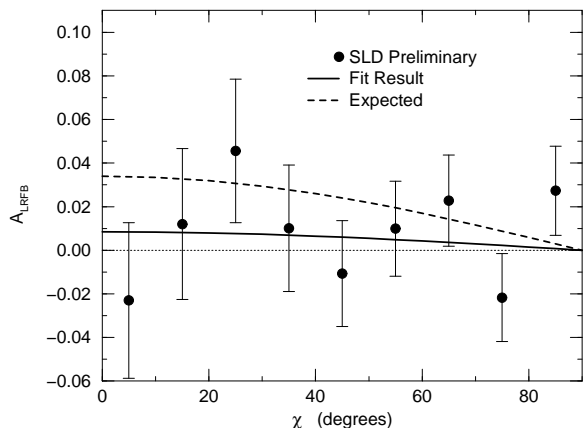}}\end{center}
\vspace {0.6cm}
\caption{ \baselineskip=12pt \label{pvchi}
Left-right-forward-backward asymmetry of the angle $\chi$ in 3-jet $Z^0$ decays.
The dashed and solid lines are the QCD prediction and the result of a fit,
respectively.}
\end{figure}

We then tag one of the two lower energy jets as the gluon jet:
if jet 2 has $n_{sig}^{(3)}=0$ and jet 3 has $n_{sig}^{(3)}>0$,
then jet 2 is tagged as
the gluon; otherwise jet 3 is tagged as the gluon.
We construct the angle $\chi$, and $A_{LRFB}^{\chi}$ is
shown as a function of $\chi$ in fig.~\ref{pvchi}.
Here we expect only a small deviation from zero as indicated by the dashed
line on fig.~\ref{pvchi}.
Our measurement is consistent with the prediction, as well as with zero.
A fit yields
$A_{\chi}\! =\! -0.014 \!\pm\! 0.035\! \pm\! 0.002$
(Preliminary), to be compared with an expectation of $-$0.064.

\section{Symmetry Tests in $Z^0\rightarrow b\bar{b}g$ Events}

Using these fully tagged events, we can construct observables that are
formally odd under time reversal and/or CP reversal.
For example, the triple product 
$\cos\omega^+\! = \!\vec{\sigma}_Z\! \cdot \!(\hat{p}_1 \!\times \!\hat{p}_2)$,
formed from the directions of the $Z^0$ polarization $\vec{\sigma}_Z$ and the
highest- and second highest-energy jets, is $T_N$-odd and CP-even.
Since the true time reversed experiment is not performed, this
quantity could have a nonzero $A_{LRFB}$,
and we have previously set a limit~\cite{evsel} using events of all flavors.
A calculation \cite{lance} including Standard Model final state interactions
predicts that $A_{LRFB}^{\omega^+}$ is largest for $b\bar{b}g$
events, but is only $\sim$10$^{-5}$.
The fully flavor-ordered triple product 
$\cos\omega^- \! =\! \vec{\sigma}_Z \!\cdot\!
(\hat{p}_q \!\times\! \hat{p}_{\bar{q}})$
is both $T_N$-odd and CP-odd.

\begin{figure}
\vspace {-0.2cm}
   \epsfxsize=4.1in
   \begin{center}\mbox{\epsffile{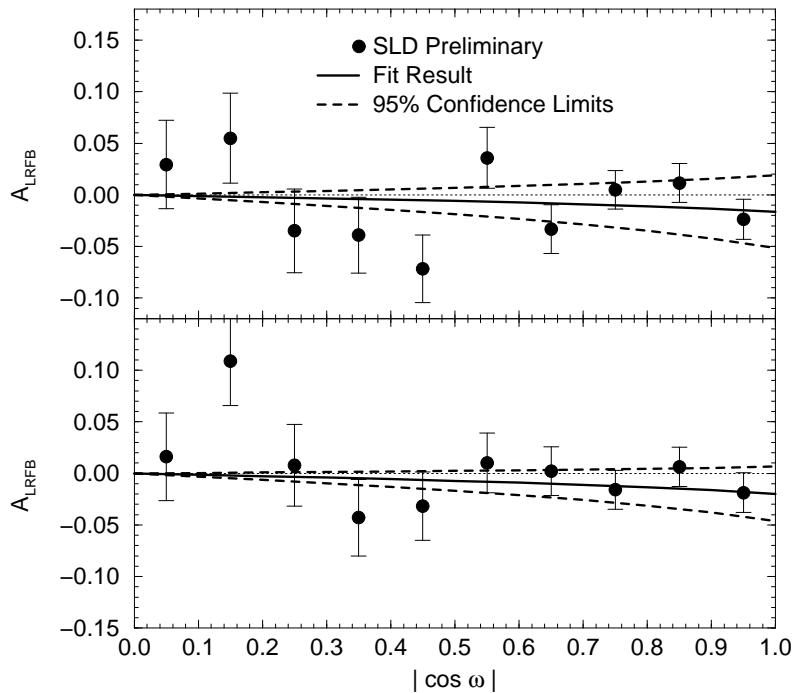}}\end{center}
\vspace {1.2cm}
\caption{ \baselineskip=12pt \label{tcp}
Left-right-forward-backward asymmetries of the energy- (top) and
flavor-ordered (bottom) triple product.
The solid (dashed) lines represent fits to the data (95\% confidence limits).
 }
\end{figure}

Our measured $A_{LRFB}^{\omega^+}$ and $A_{LRFB}^{\omega^-}$ are
shown in fig.~\ref{tcp}.
They are consistent with zero at all $|\cos\omega|$.
Fits to the data yield 95\% C.L. limits on any $T_N$-violating and
CP-conserving or CP-violating asymmetries of $-0.038\!<\!A^+_T\!<\!0.014$ or
$-0.077\!<\!A^-_T\!<\!0.011$, respectively.

\section{Gluon Splitting into a $b\bar{b}$ Pair}

Candidate events containing a gluon splitting into a $b\bar{b}$ pair, 
$Z^0 \!\rightarrow \! q\bar{q}g \!\rightarrow \!q\bar{q}b\bar{b}$,
where the initial $q\bar{q}$ can be any flavor, are required to have 4 jets
(Durham algorithm, $y_{cut}=0.008$).
A secondary vertex is required in each of the two jets with the smallest
opening angle in the event, yielding 314 events.
This sample is dominated by background, primarily from
$Z^0 \!\rightarrow \! b\bar{b}g(g)$ events and events with a gluon splitting
into a $c\bar{c}$ pair.

A large component of the former background is $Z^0\!\rightarrow \! b\bar{b}g$
events
in which the $b$ or $\bar{b}$ jet is split into two jets by the jetfinder,
and two distinct vertices from the {\it same} $B$-hadron decay are found.
Since the small beam spot allows the vertex flight directions to be
measured precisely, and the angle between the two flight directions from this
background source tends to be small, it is suppressed by a cut on this angle.

\begin{figure}
   \epsfxsize=3.9in
   \begin{center}\mbox{\epsffile{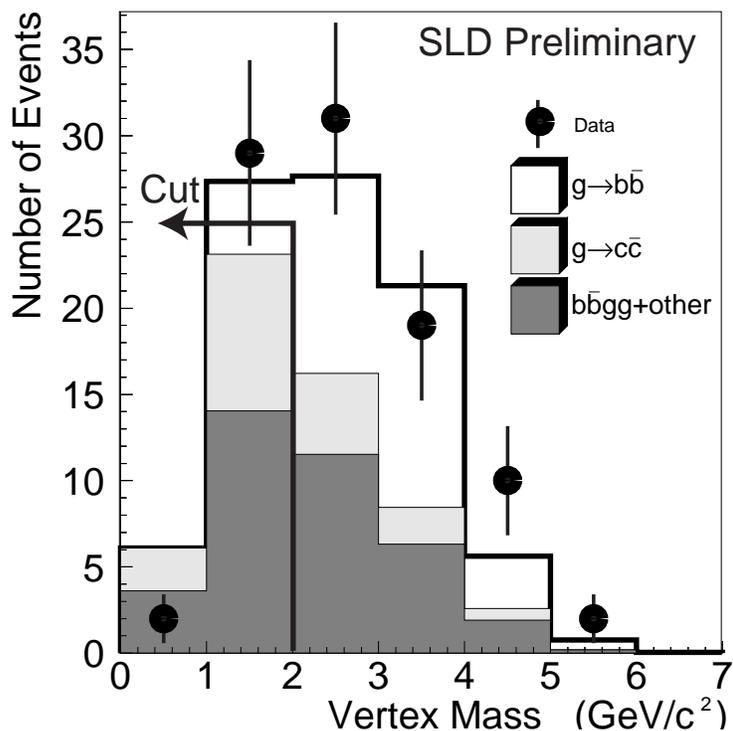}}\end{center}
\vspace {1.0cm}
\caption{ \baselineskip=12pt \label{mmax}
Distribution of the larger of the two vertex masses in candidate gluon splitting
events after all other cuts (dots).
The backgrounds expected from the simulation are indicated.
 }
\end{figure}

Cuts are also made \cite{confgbb}
on the sum of the energies of the two jets, the angle between
the plane formed by the two selected jets and that formed by
the other two jets in the event, and the larger of the vertex masses.
The distribution of the latter quantity is
shown in fig.~\ref{mmax} after all other cuts.  A clear excess of
events is visible over the expected background for masses above 2 GeV/c$^2$.
A cut at this value keeps 62 events, with an estimated background of
27.6$\pm$1.2 events.  Using this and the estimated efficiency for selecting
$g\!\rightarrow \! b\bar{b}$ splittings of 3.9\% yields a measured fraction of
hadronic events containing such a splitting of
\[
g_{b\bar{b}} = 0.0031 \pm 0.0007 \; (stat.) \pm 0.0006 \; (syst.)
\]
\noindent
(Preliminary).
The systematic error is dominated by Monte Carlo statistics.
The result is consistent with and complementary to previous measurements;
in particular it is relatively insensitive to the modelling of the gluon
splitting process, due to the excellent efficiency for finding vertices from
low-energy $B$ hadrons.

\section{The $B$ Hadron Energy Spectrum}

Any secondary vertex in either thrust hemisphere of an event that has
$m>2$ GeV/c$^2$ is considered as a candidate $B$-hadron vertex.
Its flight direction is taken to be along the line joining the
IP and the vertex position.
The four-vector sum of the tracks in the vertex (assigned the charged pion mass)
is calculated, and the momentum component $P_t$ transverse to the flight
direction is equated with the transverse component of the ``missing" momentum.
At this point two quantities are still needed in order to
determine the energy $E_B$ of the $B$ hadron, the missing mass $M_0$ and 
momentum along the flight direction.
Assuming a $B$ hadron mass of $M_B$ eliminates one of these unknowns,
and also allows an upper limit to be calculated on $M_0$:
\[
M^2_{0max} = M^2_B - 2M_B \sqrt{M^2_{chg} + P^2_t} + M^2_{chg},
\]
\noindent
where $M_{chg}$ is the mass of the set of tracks in the vertex.
Using $M_B=5.28$ GeV/c$^2$, equating $M^2_0$ with $M^2_{0max}$ and solving for
$E_B$ provides a good estimate of the true $E_B$
for the mixture of $B$ species produced in $Z^0$ decays.
As expected, the simulated energy resolution is best for vertices
with small $M^2_{0max}$, approaching 6\% as $M^2_{0max} \rightarrow 0$.
It does not degrade rapidly with increasing $M^2_{0max}$
due to the strong tendency for the true $M_0$ to
cluster near the maximum value.

\begin{figure}[t]
   \epsfxsize=5.9in
   \begin{center}\mbox{\epsffile{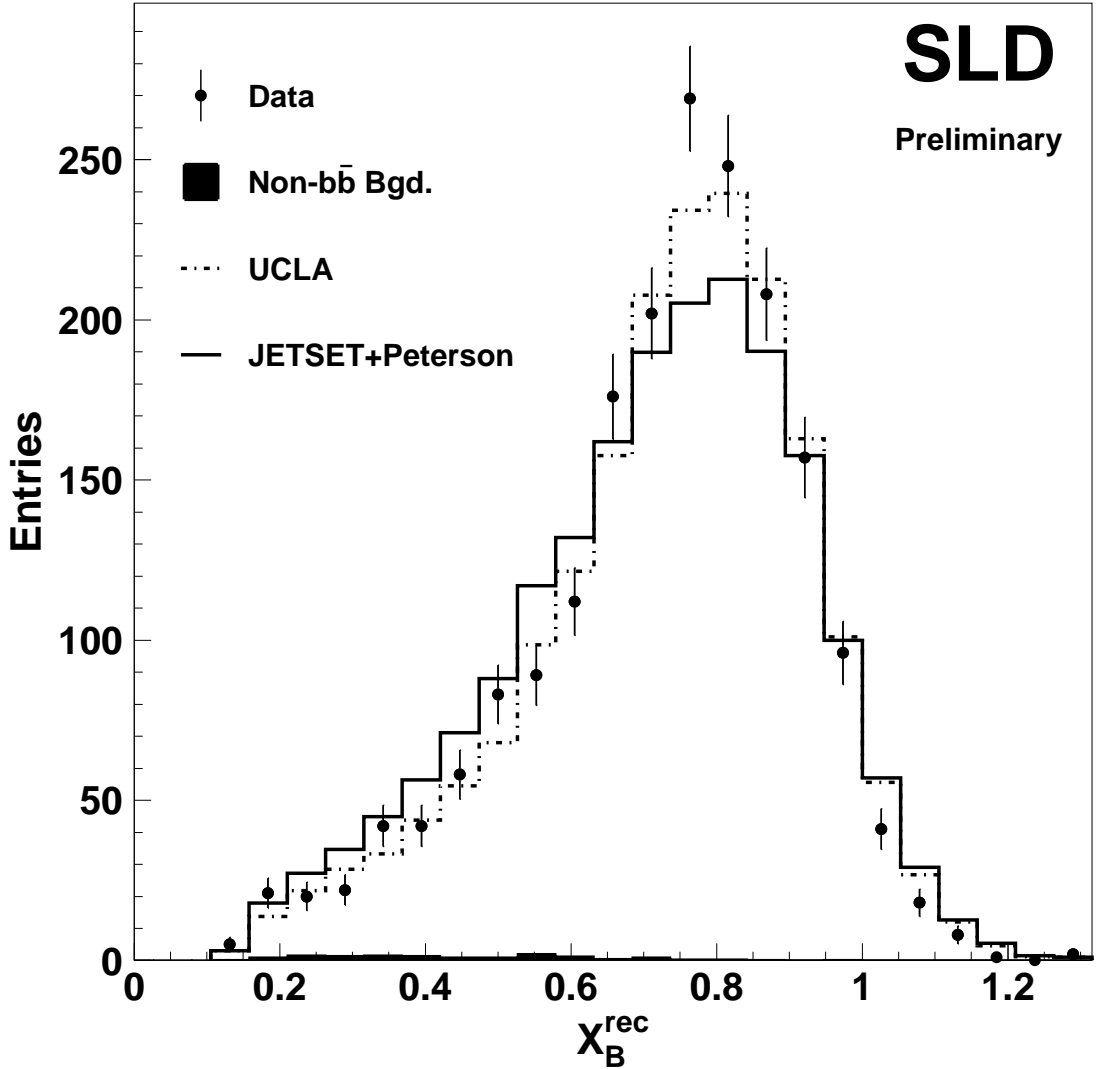}}\end{center}
\vspace {1.9cm}
\caption{ \baselineskip=12pt \label{ebfig}
Uncorrected distribution of reconstructed $B$-hadron energies (dots).
The solid (dot-dashed) histogram is the prediction of the JETSET$+$Peterson
(UCLA) simulation.}
\end{figure}

A cut is placed on $M^2_{0max}$ that depends on the measured $E_B$ in such a
way that the simulated efficiency for selecting $B$ vertices
is roughly independent of energy;
it is 4\% on average and is above 3\% for $E_B>8$ GeV.
A sample of 1938 vertices is selected from one-third of the hadronic event
sample, with an estimated $B$-hadron purity of 99.5\%.
The simulated energy resolution is 10\% on average, roughly independent
of $E_B$.
The raw distribution of the scaled energy $x_B=E_B/E_{beam}$ is shown in
fig.~\ref{ebfig}, and covers the entire kinematic range from the
$B$-hadron mass ($x_B \approx 0.12$) to the beam energy.

Also shown in fig.~\ref{ebfig} is the prediction of our simulation, generated
using the JETSET program with the Peterson fragmentation option and
$\epsilon_b=0.006$.
The predicted peak position is consistent with that of the data, but the width
is significantly larger than that of the data.

The correction of these data to obtain the true $x_B$ 
distribution depends on the form assumed for the true distribution,
due to the rapid variation of the distribution on the scale of
the bin size.
We therefore test several hypothesized shapes by weighting our simulated
events to reproduce a given function at the generator level, and comparing the
corresponding detector level prediction with the data in fig.~\ref{ebfig} using
a binned $\chi^2$.

We first consider a number of heavy-hadron fragmentation models within the
context of the JETSET simulation, by generating samples of 0.5 million
$Z^0\!\rightarrow \! b\bar{b}$ events for each of several values of the free
parameters of each model.
Minimizing $\chi^2$ with respect to the parameter $\epsilon_b$ of the Peterson
function results in a fitted value of $\epsilon_b=0.006$, equal to our nominal
value; however the $\chi^2$ of 62 for 16 degrees of freedom indicates that
this model is inconsistent with the data.
We also exclude \cite{confxb} the models of Braaten et.~al, and Collins and
Spiller, whereas those of the Lund group, Bowler, and Kartvelishvili are able
to describe the data.
In addition we test the UCLA and HERWIG fragmentation models, in which there
are no explicit free parameters governing $B$ hadron production.  The prediction
of the UCLA model (see fig.~\ref{ebfig}) is consistent with the
data, but that of the HERWIG model is not.

We also test several ad hoc functional forms $f(x_B,\vec{\lambda})$ of the
observable $x_B$ itself, by minimizing $\chi^2$ with respect to the
parameter(s) $\vec{\lambda}$.
Most of our test functions \cite{confxb} are not able to describe the
data for any values of their parameters.
Four functions, the Peterson function, two generalizations thereof, and a sixth
order polynomial, are found to describe the data.

\begin{figure}[t]
   \epsfxsize=5.4in
   \begin{center}\mbox{\epsffile{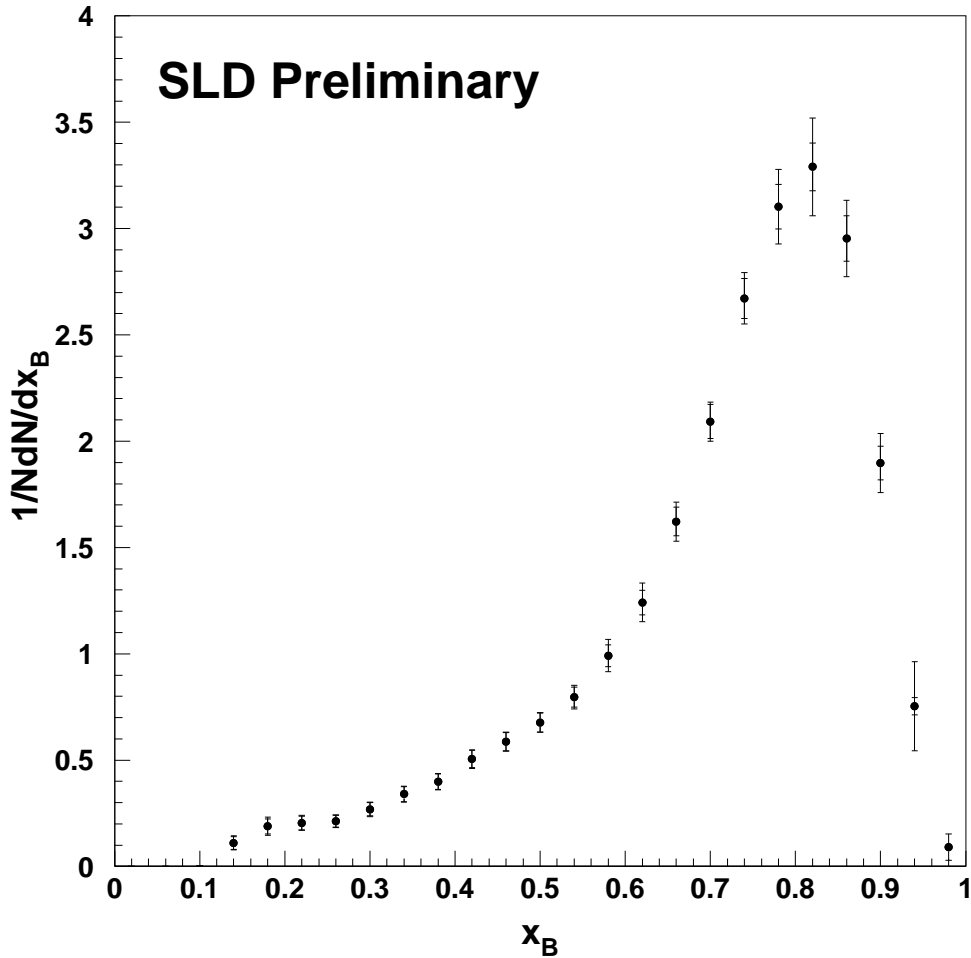}}\end{center}
\vspace {1.5cm}
\caption{ \baselineskip=12pt \label{ebcorr}
Corrected distribution of $B$-hadron energies averaged over the eight
acceptable shapes.
The outer error bars include the rms deviation among these shapes and provide
an envelope for the true shape of the distribution.}
\end{figure}

We subtract the estimated background and unfold the data to obtain the true
distribution $D^{corr}_i = \Sigma_k M_{ik} D^{meas}_k / \epsilon_i$,
where the efficiency $\vec{\epsilon}$ and the unfolding matrix $\bf M$ are
calculated from the simulation using in turn each of the four fragmentation
models and four functional forms that are consistent with the data.
The model dependence of the procedure is thus made explicit, and is
visible in fig.~\ref{ebcorr}, where in each bin $i$ we show the average of
the eight $D^{corr}_i$, and the error bar includes their rms deviation,
which is substantial at high $x_B$.
The corrected distribution is, by construction, smoother than the measured
distribution, and the error bars provide a 1$\sigma$ envelope within which any
acceptable prediction must fall.

From these eight forms we extract a measurement of the mean value of the
scaled energy,
\[
\left< x_B \right> = 0.714 \pm 0.005 (stat.) \pm 0.007 (syst.) \pm 0.002 (rms)
\]
\noindent
(Preliminary).
This is the most precise of the world's measurements that take the shape
dependence into account, and this uncertainty is small
since we are able to exclude a wide range of shapes.

\begin{figure}
   \epsfxsize=3.2in
   \begin{center}\mbox{\epsffile{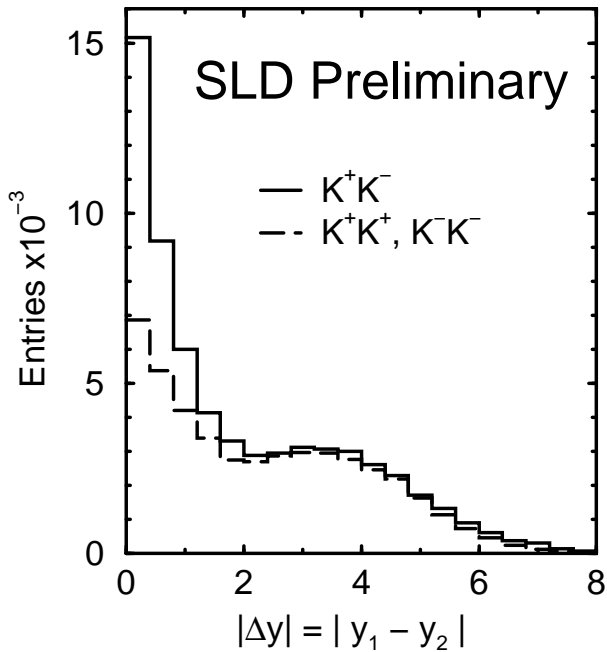}}\end{center}
\vspace {1.1cm}
\caption{ \baselineskip=12pt \label{dykk}
Distributions of the absolute rapidity difference $|\Delta y|$ for
opposite-charge (histogram) and
same-charge (dashed histrogram) pairs of identified charged kaons in hadronic
$Z^0$ decays.
 }
\end{figure}

\section{Rapidity Correlations}

\begin{figure}[t]
   \epsfxsize=3.7in
   \begin{center}\mbox{\epsffile{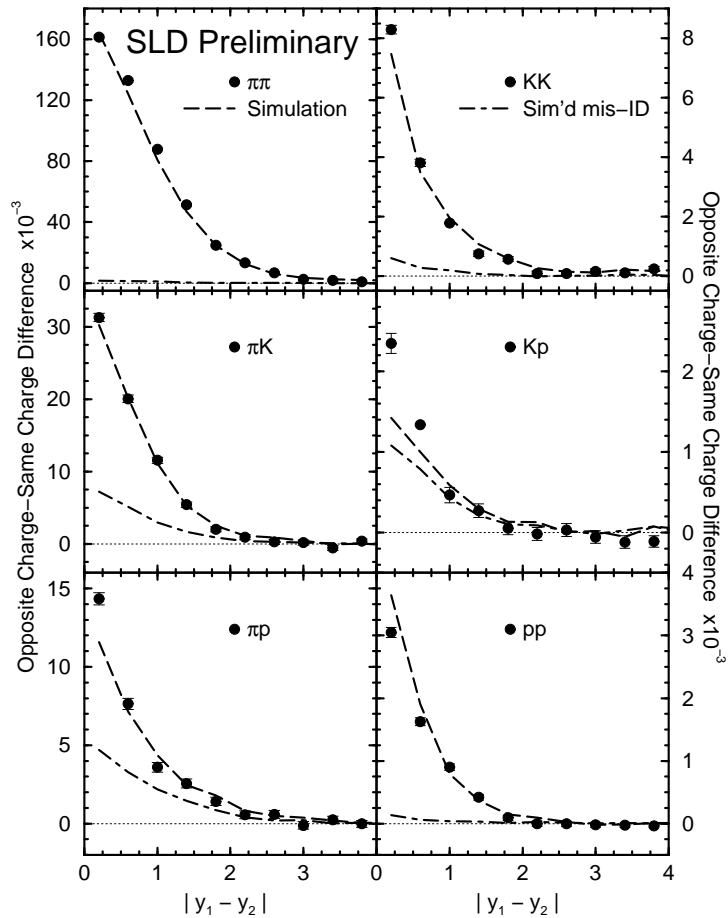}}\end{center}
\vspace {1.0cm}
\caption{ \baselineskip=12pt \label{ddys}
Differences between the $|\Delta y|$ distributions for opposite- and
same-charge pairs of identified $\pi^\pm$, $K^\pm$, p/$\bar{\rm p}$ in hadronic
$Z^0$ decays (dots).
The dashed (dot-dashed) lines indicate the simulated differences (contributions
from pairs with a misidentified track).
 }
\end{figure}

For this study \cite{confcorl} we use the entire sample of hadronic events,
as well as subsamples
tagged as primary light-($uds$), $c$-, and $b$-flavor, having purities of
88\%, 39\%, and 93\%, respectively.
Charged tracks identified as $\pi^\pm$, $K^\pm$ or p/$\bar{\rm p}$ in the CRID
are considered, and their rapidities
$y=0.5\ln ((E+p_{\parallel})/(E-p_{\parallel}))$ are calculated using their
measured energies and components of momentum along the event thrust axis
$p_{\parallel}$.
For each pair of identified tracks in an event the absolute
value of the difference between their rapidities $|\Delta y|=|y_1 - y_2|$ is
considered.
Figure \ref{dykk} shows the distributions of $|\Delta y|$ for identified
$K^+K^-$ pairs and for $K^+K^+$/$K^-K^-$ pairs.  The latter are assumed to be
uncorrelated, and the excess of the former at low values of $|\Delta y|$ is
interpreted as a short-range correlation, indicating that
strangeness conservation is local in the jet fragmentation process.
This is also the case for baryon
number and electric charge, as is well known and visible in fig.~\ref{ddys},
which shows the difference between the opposite-charge and same-charge
distributions for all six pair combinations.

Also visible in fig.~\ref{ddys} are short-range correlations for all
three unlike pair combinations.  
Excellent particle identification is required to observe these over the
background from $\pi\pi$ pairs in which one of the pions is misidentified.
This is the first direct observation of a fundamental feature of the jet
fragmentation process, that electric charge can be conserved locally between
a meson and a baryon, or between a strange particle and a nonstrange particle,
and suggests charge ordering along the entire fragmentation chain.

\nopagebreak
The high statistics and wide momentum coverage allow a number of detailed
measurements.
The predictions of the JETSET model (see fig. \ref{ddys})
describe the amplitude and range of the observed correlations except for
$K$p pairs.
This is true in six bins of momentum \cite{confcorl}, and we conclude that,
within the context of the JETSET model, the correlations are scale invariant.

\begin{figure}[t]
   \epsfxsize=3.9in
   \begin{center}\mbox{\epsffile{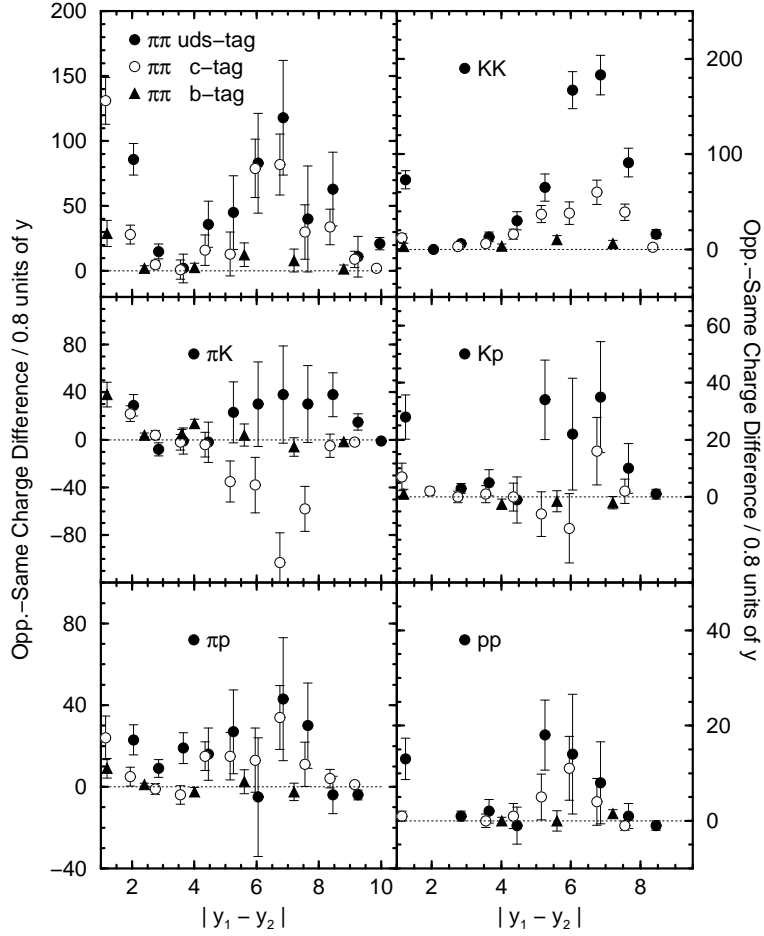}}\end{center}
\vspace {1.0cm}
\caption{ \baselineskip=12pt \label{ddyl}
Differences between the $|\Delta y|$ distributions for opposite-charge and
same-charge pairs in which both tracks have $p>9$ GeV/c, for the light-(dots),
$c$-(open circles), and $b$-tagged (triangles) samples.
 }
\end{figure}

To study long-range correlations, which are expected from leading
particle production,  it is necessary to consider high momentum tracks.
The differences
between the opposite- and same-charge $|\Delta y|$ distributions for pairs of
tracks in which both have $p>9$ GeV/c are shown for each of the three
flavor-tagged samples in fig.~\ref{ddyl}.
A very large $K^+K^-$ correlation is observed in light-flavor events,
as expected from
$s\bar{s}$ events in which the $s$ ($\bar{s}$) jet produces a leading $K^-$
($K^+$).
For all other combinations, any correlation will be diluted by
the short-range correlation -- e.g. a leading baryon will be accompanied by
a subleading antibaryon with similar $y$.
However significant
correlations are observed for all pair combinations in light flavor events,
providing new information on leading particle production.
The JETSET model predictions are consistent with these data \cite{confcorl},
except that no $\pi K$ correlation is predicted.
Flavor tagging is essential for these measurements; decays of leading charmed
hadrons contribute to several correlations,
in particular they give an anticorrelation for $\pi K$ pairs that would mask
the signal from light flavors.

We now give the rapidity a meaningful sign by using the beam polarization to
tag the quark
hemisphere in each event, with a purity of 73\%.
The thrust axis is signed to point into this hemisphere, thus signing the
rapidity such that $y>0$ ($y<0$) in the (anti)quark hemisphere.
For pairs of hadrons one can define an ordered rapidity difference;
for hadron-antihadron pairs we define
$\Delta y^{+-} = y_+ - y_-$ as the rapidity of the positively charged track
minus that of the negative track.

\begin{figure}[t]
   \epsfxsize=4.0in
   \begin{center}\mbox{\epsffile{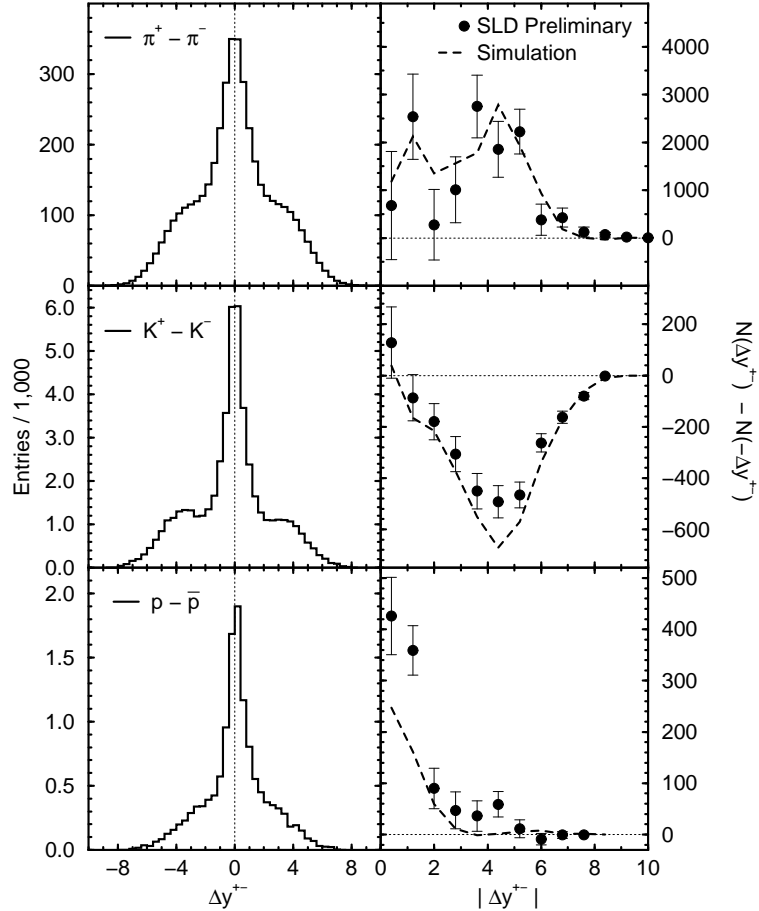}}\end{center}
\vspace {1.0cm}
\caption{ \baselineskip=12pt \label{dsy}
Distributions (left) of the ordered signed rapidity difference $\Delta y^{+-}$,
and differences (right) between the positive and negative sides of each
distribution.
The dashed lines indicate the predictions of the JETSET simulation.
 }
\end{figure}

A distribution of $\Delta y^{+-}$ can be studied independent of
the same-charge pairs by considering asymmetries between the
positive and negative sides of the distribution.
In fig.~\ref{dsy} we show the distributions of $\Delta y^{+-}$, along with the
differences between the two sides, for $\pi^+\pi^-$,
$K^+K^-$ and p$\bar{\rm p}$ pairs, which show a number of nonzero asymmetries.
At long-range ($|\Delta y^{+-}| > 3)$, such deviations are expected from leading
particle production.  A large negative difference is observed for $K^+K^-$
pairs, as expected; a corresponding small positive difference is visible
for p$\bar{\rm p}$ pairs, but a small negative difference for
$\pi^+\pi^-$ pairs is not.
Flavor tagging again proves beneficial:  the positive differences
at all $|\Delta y^{+-}|$ are present only in the $c$-tagged sample,
and are explained by the leading charmed hadrons;
a small negative difference is observed in the light-flavor
sample \cite{confcorl}.

The large positive difference observed for p$\bar{\rm p}$ pairs at low
$|\Delta y^{+-}|$ is the first direct observation of another fundamental
feature of jet fragmentation, namely the ordering of baryon number along the
quark-antiquark axis.
That is, the proton in a correlated
proton-antiproton pair `knows' and prefers the quark direction over the
antiquark direction.
This excess is observed at all proton momenta so cannot be attributed simply
to leading baryons.
We have searched for similar signals for strangeness and charge ordering in the
$K^+K^-$ and $\pi^+\pi^-$ samples, respectively, by isolating the light flavors
and considering a variety of momentum bins.
However no significant effects are observed, possibly due to background from
leading kaons and/or dilution due to resonance decays.

\section{Conclusions}

We use the excellent SLD vertexing and particle identificaton, and
the high SLC $e^-$ beam polarization to make several new
tests of QCD in the two areas of event structure and jet fragmentation.
We find the $Z^0\!\rightarrow \! b\bar{b}g$ rate to be sensitive to the
$b$-quark mass, providing a promising new way to measure $m_b$.
Using 3-jet final states in which jets are tagged as $b$,
$\bar{b}$ or gluon jets:
we measure the gluon energy spectrum over its full range,
confirming the prediction of QCD and setting limits on anomalous
chromomagnetic couplings;
we find the parity violation in $Z^0\!\rightarrow \! b\bar{b}g$ decays to be
consistent with electroweak theory plus QCD;
and we perform new tests of T- and CP-conservation in strong interactions.
Using 4-jet final states in which the two most collinear jets are tagged as
$b$/$\bar{b}$, we measure the rate of gluon splitting into a $b\bar{b}$ pair in
hadronic $Z^0$ decays, 
$g_{b\bar{b}} = 0.0031 \pm 0.0007 (stat.) \pm 0.0006 (syst.)$ (Preliminary).

A new, inclusive technique for measuring the energies of individual $B$ hadrons
uses information only from the charged tracks attached to a secondary vertex
to give good efficiency and resolution at all energies.
The full $B$-hadron energy distribution is measured precisely, allowing us to
test a number of model predictions and ad hoc functional forms.
We exclude a wide range of forms, constrain the shape tightly,
and obtain a precise measurement of the average
scaled energy,
$\left< x_B \right> = 0.714 \pm 0.005 (stat.) \pm 0.007 (syst.) \pm 0.002$ ({\it shape})
(Preliminary).

Considering pairs of identified $\pi^\pm$, $K^\pm$ and p/$\bar{\rm p}$,
we confirm that the conservation of
quantum numbers
is local in the jet fragmentation process, and observe local charge conservation
between mesons and baryons and between strange and nonstrange particles.
Long range correlations are also observed between all these pair combinations,
providing new information on leading particle production.
The first study of ordered correlations in signed rapidity provides
additional new information on fragmentation, including the first
direct observation of baryon number ordering along the quark-antiquark axis.

%


\section*{$^{**}$List of Authors} 

%
%
%
\begin{center}
\def\iADEL{$^{(1)}$}
\def\iAOMORI{$^{(2)}$}
\def\iBOLO{$^{(3)}$}
\def\iBRI{$^{(4)}$}
\def\iBRUN{$^{(5)}$}
\def\iBU{$^{(6)}$}
\def\iCINC{$^{(7)}$}
\def\iCOLO{$^{(8)}$}
\def\iCOLU{$^{(9)}$}
\def\iCSU{$^{(10)}$}
\def\iFERR{$^{(11)}$}
\def\iFRAS{$^{(12)}$}
\def\iILLI{$^{(13)}$}
\def\iJHU{$^{(14)}$}
\def\iLBL{$^{(15)}$}
\def\iLTU{$^{(16)}$}
\def\iMASS{$^{(17)}$}
\def\iMISSI{$^{(18)}$}
\def\iMIT{$^{(19)}$}
\def\iMOSCOW{$^{(20)}$}
\def\iNAGO{$^{(21)}$}
\def\iOREG{$^{(22)}$}
\def\iOXF{$^{(23)}$}
\def\iPADO{$^{(24)}$}
\def\iPERU{$^{(25)}$}
\def\iPISA{$^{(26)}$}
\def\iRAL{$^{(27)}$}
\def\iRUTG{$^{(28)}$}
\def\iSLAC{$^{(29)}$}
\def\iSOGA{$^{(30)}$}
\def\iSOONG{$^{(31)}$}
\def\iTENN{$^{(32)}$}
\def\iTOHO{$^{(33)}$}
\def\iUCSB{$^{(34)}$}
\def\iUCSC{$^{(35)}$}
\def\iUVIC{$^{(36)}$}
\def\iVAND{$^{(37)}$}
\def\iWASH{$^{(38)}$}
\def\iWISC{$^{(39)}$}
\def\iYALE{$^{(40)}$}

  \baselineskip=.75\baselineskip  
\mbox{Kenji  Abe\unskip,\iNAGO}
\mbox{Koya Abe\unskip,\iTOHO}
\mbox{T. Abe\unskip,\iSLAC}
\mbox{I.Adam\unskip,\iSLAC}
\mbox{T.  Akagi\unskip,\iSLAC}
\mbox{N. J. Allen\unskip,\iBRUN}
\mbox{W.W. Ash\unskip,\iSLAC}
\mbox{D. Aston\unskip,\iSLAC}
\mbox{K.G. Baird\unskip,\iMASS}
\mbox{C. Baltay\unskip,\iYALE}
\mbox{H.R. Band\unskip,\iWISC}
\mbox{M.B. Barakat\unskip,\iLTU}
\mbox{O. Bardon\unskip,\iMIT}
\mbox{T.L. Barklow\unskip,\iSLAC}
\mbox{G. L. Bashindzhagyan\unskip,\iMOSCOW}
\mbox{J.M. Bauer\unskip,\iMISSI}
\mbox{G. Bellodi\unskip,\iOXF}
\mbox{R. Ben-David\unskip,\iYALE}
\mbox{A.C. Benvenuti\unskip,\iBOLO}
\mbox{G.M. Bilei\unskip,\iPERU}
\mbox{D. Bisello\unskip,\iPADO}
\mbox{G. Blaylock\unskip,\iMASS}
\mbox{J.R. Bogart\unskip,\iSLAC}
\mbox{G.R. Bower\unskip,\iSLAC}
\mbox{J. E. Brau\unskip,\iOREG}
\mbox{M. Breidenbach\unskip,\iSLAC}
\mbox{W.M. Bugg\unskip,\iTENN}
\mbox{D. Burke\unskip,\iSLAC}
\mbox{T.H. Burnett\unskip,\iWASH}
\mbox{P.N. Burrows\unskip,\iOXF}
\mbox{A. Calcaterra\unskip,\iFRAS}
\mbox{D. Calloway\unskip,\iSLAC}
\mbox{B. Camanzi\unskip,\iFERR}
\mbox{M. Carpinelli\unskip,\iPISA}
\mbox{R. Cassell\unskip,\iSLAC}
\mbox{R. Castaldi\unskip,\iPISA}
\mbox{A. Castro\unskip,\iPADO}
\mbox{M. Cavalli-Sforza\unskip,\iUCSC}
\mbox{A. Chou\unskip,\iSLAC}
\mbox{E. Church\unskip,\iWASH}
\mbox{H.O. Cohn\unskip,\iTENN}
\mbox{J.A. Coller\unskip,\iBU}
\mbox{M.R. Convery\unskip,\iSLAC}
\mbox{V. Cook\unskip,\iWASH}
\mbox{R. Cotton\unskip,\iBRUN}
\mbox{R.F. Cowan\unskip,\iMIT}
\mbox{D.G. Coyne\unskip,\iUCSC}
\mbox{G. Crawford\unskip,\iSLAC}
\mbox{C.J.S. Damerell\unskip,\iRAL}
\mbox{M. N. Danielson\unskip,\iCOLO}
\mbox{M. Daoudi\unskip,\iSLAC}
\mbox{N. de Groot\unskip,\iBRI}
\mbox{R. Dell'Orso\unskip,\iPERU}
\mbox{P.J. Dervan\unskip,\iBRUN}
\mbox{R. de Sangro\unskip,\iFRAS}
\mbox{M. Dima\unskip,\iCSU}
\mbox{A. D'Oliveira\unskip,\iCINC}
\mbox{D.N. Dong\unskip,\iMIT}
\mbox{M. Doser\unskip,\iSLAC}
\mbox{R. Dubois\unskip,\iSLAC}
\mbox{B.I. Eisenstein\unskip,\iILLI}
\mbox{V. Eschenburg\unskip,\iMISSI}
\mbox{E. Etzion\unskip,\iWISC}
\mbox{S. Fahey\unskip,\iCOLO}
\mbox{D. Falciai\unskip,\iFRAS}
\mbox{C. Fan\unskip,\iCOLO}
\mbox{J.P. Fernandez\unskip,\iUCSC}
\mbox{M.J. Fero\unskip,\iMIT}
\mbox{K.Flood\unskip,\iMASS}
\mbox{R. Frey\unskip,\iOREG}
\mbox{J. Gifford\unskip,\iUVIC}
\mbox{T. Gillman\unskip,\iRAL}
\mbox{G. Gladding\unskip,\iILLI}
\mbox{S. Gonzalez\unskip,\iMIT}
\mbox{E. R. Goodman\unskip,\iCOLO}
\mbox{E.L. Hart\unskip,\iTENN}
\mbox{J.L. Harton\unskip,\iCSU}
\mbox{A. Hasan\unskip,\iBRUN}
\mbox{K. Hasuko\unskip,\iTOHO}
\mbox{S. J. Hedges\unskip,\iBU}
\mbox{S.S. Hertzbach\unskip,\iMASS}
\mbox{M.D. Hildreth\unskip,\iSLAC}
\mbox{J. Huber\unskip,\iOREG}
\mbox{M.E. Huffer\unskip,\iSLAC}
\mbox{E.W. Hughes\unskip,\iSLAC}
\mbox{X.Huynh\unskip,\iSLAC}
\mbox{H. Hwang\unskip,\iOREG}
\mbox{M. Iwasaki\unskip,\iOREG}
\mbox{D. J. Jackson\unskip,\iRAL}
\mbox{P. Jacques\unskip,\iRUTG}
\mbox{J.A. Jaros\unskip,\iSLAC}
\mbox{Z.Y. Jiang\unskip,\iSLAC}
\mbox{A.S. Johnson\unskip,\iSLAC}
\mbox{J.R. Johnson\unskip,\iWISC}
\mbox{R.A. Johnson\unskip,\iCINC}
\mbox{T. Junk\unskip,\iSLAC}
\mbox{R. Kajikawa\unskip,\iNAGO}
\mbox{M. Kalelkar\unskip,\iRUTG}
\mbox{Y. Kamyshkov\unskip,\iTENN}
\mbox{H.J. Kang\unskip,\iRUTG}
\mbox{I. Karliner\unskip,\iILLI}
\mbox{H. Kawahara\unskip,\iSLAC}
\mbox{Y. D. Kim\unskip,\iSOGA}
\mbox{M.E. King\unskip,\iSLAC}
\mbox{R. King\unskip,\iSLAC}
\mbox{R.R. Kofler\unskip,\iMASS}
\mbox{N.M. Krishna\unskip,\iCOLO}
\mbox{R.S. Kroeger\unskip,\iMISSI}
\mbox{M. Langston\unskip,\iOREG}
\mbox{A. Lath\unskip,\iMIT}
\mbox{D.W.G. Leith\unskip,\iSLAC}
\mbox{V. Lia\unskip,\iMIT}
\mbox{C.Lin\unskip,\iMASS}
\mbox{M.X. Liu\unskip,\iYALE}
\mbox{X. Liu\unskip,\iUCSC}
\mbox{M. Loreti\unskip,\iPADO}
\mbox{A. Lu\unskip,\iUCSB}
\mbox{H.L. Lynch\unskip,\iSLAC}
\mbox{J. Ma\unskip,\iWASH}
\mbox{G. Mancinelli\unskip,\iRUTG}
\mbox{S. Manly\unskip,\iYALE}
\mbox{G. Mantovani\unskip,\iPERU}
\mbox{T.W. Markiewicz\unskip,\iSLAC}
\mbox{T. Maruyama\unskip,\iSLAC}
\mbox{H. Masuda\unskip,\iSLAC}
\mbox{E. Mazzucato\unskip,\iFERR}
\mbox{A.K. McKemey\unskip,\iBRUN}
\mbox{B.T. Meadows\unskip,\iCINC}
\mbox{G. Menegatti\unskip,\iFERR}
\mbox{R. Messner\unskip,\iSLAC}
\mbox{P.M. Mockett\unskip,\iWASH}
\mbox{K.C. Moffeit\unskip,\iSLAC}
\mbox{T.B. Moore\unskip,\iYALE}
\mbox{M.Morii\unskip,\iSLAC}
\mbox{D. Muller\unskip,\iSLAC}
\mbox{V.Murzin\unskip,\iMOSCOW}
\mbox{T. Nagamine\unskip,\iTOHO}
\mbox{S. Narita\unskip,\iTOHO}
\mbox{U. Nauenberg\unskip,\iCOLO}
\mbox{H. Neal\unskip,\iSLAC}
\mbox{M. Nussbaum\unskip,\iCINC}
\mbox{N.Oishi\unskip,\iNAGO}
\mbox{D. Onoprienko\unskip,\iTENN}
\mbox{L.S. Osborne\unskip,\iMIT}
\mbox{R.S. Panvini\unskip,\iVAND}
\mbox{C. H. Park\unskip,\iSOONG}
\mbox{T.J. Pavel\unskip,\iSLAC}
\mbox{I. Peruzzi\unskip,\iFRAS}
\mbox{M. Piccolo\unskip,\iFRAS}
\mbox{L. Piemontese\unskip,\iFERR}
\mbox{K.T. Pitts\unskip,\iOREG}
\mbox{R.J. Plano\unskip,\iRUTG}
\mbox{R. Prepost\unskip,\iWISC}
\mbox{C.Y. Prescott\unskip,\iSLAC}
\mbox{G.D. Punkar\unskip,\iSLAC}
\mbox{J. Quigley\unskip,\iMIT}
\mbox{B.N. Ratcliff\unskip,\iSLAC}
\mbox{T.W. Reeves\unskip,\iVAND}
\mbox{J. Reidy\unskip,\iMISSI}
\mbox{P.L. Reinertsen\unskip,\iUCSC}
\mbox{P.E. Rensing\unskip,\iSLAC}
\mbox{L.S. Rochester\unskip,\iSLAC}
\mbox{P.C. Rowson\unskip,\iCOLU}
\mbox{J.J. Russell\unskip,\iSLAC}
\mbox{O.H. Saxton\unskip,\iSLAC}
\mbox{T. Schalk\unskip,\iUCSC}
\mbox{R.H. Schindler\unskip,\iSLAC}
\mbox{B.A. Schumm\unskip,\iUCSC}
\mbox{J. Schwiening\unskip,\iSLAC}
\mbox{S. Sen\unskip,\iYALE}
\mbox{V.V. Serbo\unskip,\iSLAC}
\mbox{M.H. Shaevitz\unskip,\iCOLU}
\mbox{J.T. Shank\unskip,\iBU}
\mbox{G. Shapiro\unskip,\iLBL}
\mbox{D.J. Sherden\unskip,\iSLAC}
\mbox{K. D. Shmakov\unskip,\iTENN}
\mbox{C. Simopoulos\unskip,\iSLAC}
\mbox{N.B. Sinev\unskip,\iOREG}
\mbox{S.R. Smith\unskip,\iSLAC}
\mbox{M. B. Smy\unskip,\iCSU}
\mbox{J.A. Snyder\unskip,\iYALE}
\mbox{H. Staengle\unskip,\iCSU}
\mbox{A. Stahl\unskip,\iSLAC}
\mbox{P. Stamer\unskip,\iRUTG}
\mbox{H. Steiner\unskip,\iLBL}
\mbox{R. Steiner\unskip,\iADEL}
\mbox{M.G. Strauss\unskip,\iMASS}
\mbox{D. Su\unskip,\iSLAC}
\mbox{F. Suekane\unskip,\iTOHO}
\mbox{A. Sugiyama\unskip,\iNAGO}
\mbox{S. Suzuki\unskip,\iNAGO}
\mbox{M. Swartz\unskip,\iJHU}
\mbox{A. Szumilo\unskip,\iWASH}
\mbox{T. Takahashi\unskip,\iSLAC}
\mbox{F.E. Taylor\unskip,\iMIT}
\mbox{J. Thom\unskip,\iSLAC}
\mbox{E. Torrence\unskip,\iMIT}
\mbox{N. K. Toumbas\unskip,\iSLAC}
\mbox{T. Usher\unskip,\iSLAC}
\mbox{C. Vannini\unskip,\iPISA}
\mbox{J. Va'vra\unskip,\iSLAC}
\mbox{E. Vella\unskip,\iSLAC}
\mbox{J.P. Venuti\unskip,\iVAND}
\mbox{R. Verdier\unskip,\iMIT}
\mbox{P.G. Verdini\unskip,\iPISA}
\mbox{D. L. Wagner\unskip,\iCOLO}
\mbox{S.R. Wagner\unskip,\iSLAC}
\mbox{A.P. Waite\unskip,\iSLAC}
\mbox{S. Walston\unskip,\iOREG}
\mbox{J.Wang\unskip,\iSLAC}
\mbox{S.J. Watts\unskip,\iBRUN}
\mbox{A.W. Weidemann\unskip,\iTENN}
\mbox{E. R. Weiss\unskip,\iWASH}
\mbox{J.S. Whitaker\unskip,\iBU}
\mbox{S.L. White\unskip,\iTENN}
\mbox{F.J. Wickens\unskip,\iRAL}
\mbox{B. Williams\unskip,\iCOLO}
\mbox{D.C. Williams\unskip,\iMIT}
\mbox{S.H. Williams\unskip,\iSLAC}
\mbox{S. Willocq\unskip,\iMASS}
\mbox{R.J. Wilson\unskip,\iCSU}
\mbox{W.J. Wisniewski\unskip,\iSLAC}
\mbox{J. L. Wittlin\unskip,\iMASS}
\mbox{M. Woods\unskip,\iSLAC}
\mbox{G.B. Word\unskip,\iVAND}
\mbox{T.R. Wright\unskip,\iWISC}
\mbox{J. Wyss\unskip,\iPADO}
\mbox{R.K. Yamamoto\unskip,\iMIT}
\mbox{J.M. Yamartino\unskip,\iMIT}
\mbox{X. Yang\unskip,\iOREG}
\mbox{J. Yashima\unskip,\iTOHO}
\mbox{S.J. Yellin\unskip,\iUCSB}
\mbox{C.C. Young\unskip,\iSLAC}
\mbox{H. Yuta\unskip,\iAOMORI}
\mbox{G. Zapalac\unskip,\iWISC}
\mbox{R.W. Zdarko\unskip,\iSLAC}
\mbox{J. Zhou\unskip.\iOREG}

\it
  \vskip \baselineskip                   
  \vskip \baselineskip        
  \baselineskip=.75\baselineskip   
\iADEL
  Adelphi University, Garden City, New York 11530, \break
\iAOMORI
  Aomori University, Aomori , 030 Japan, \break
\iBOLO
  INFN Sezione di Bologna, I-40126, Bologna Italy, \break
\iBRI
  University of Bristol, Bristol, U.K., \break
\iBRUN
  Brunel University, Uxbridge, Middlesex, UB8 3PH United Kingdom, \break
\iBU
  Boston University, Boston, Massachusetts 02215, \break
\iCINC
  University of Cincinnati, Cincinnati, Ohio 45221, \break
\iCOLO
  University of Colorado, Boulder, Colorado 80309, \break
\iCOLU
  Columbia University, New York, New York 10533, \break
\iCSU
  Colorado State University, Ft. Collins, Colorado 80523, \break
\iFERR
  INFN Sezione di Ferrara and Universita di Ferrara, I-44100 Ferrara, Italy, \break
\iFRAS
  INFN Lab. Nazionali di Frascati, I-00044 Frascati, Italy, \break
\iILLI
  University of Illinois, Urbana, Illinois 61801, \break
\iJHU
  Johns Hopkins University, Baltimore, MD 21218-2686, \break
\iLBL
  Lawrence Berkeley Laboratory, University of California, Berkeley, California 94720, \break
\iLTU
  Louisiana Technical University - Ruston,LA 71272, \break
\iMASS
  University of Massachusetts, Amherst, Massachusetts 01003, \break
\iMISSI
  University of Mississippi, University, Mississippi 38677, \break
\iMIT
  Massachusetts Institute of Technology, Cambridge, Massachusetts 02139, \break
\iMOSCOW
  Institute of Nuclear Physics, Moscow State University, 119899, Moscow Russia, \break
\iNAGO
  Nagoya University, Chikusa-ku, Nagoya 464 Japan, \break
\iOREG
  University of Oregon, Eugene, Oregon 97403, \break
\iOXF
  Oxford University, Oxford, OX1 3RH, United Kingdom, \break
\iPADO
  INFN Sezione di Padova and Universita di Padova I-35100, Padova, Italy, \break
\iPERU
  INFN Sezione di Perugia and Universita di Perugia, I-06100 Perugia, Italy, \break
\iPISA
  INFN Sezione di Pisa and Universita di Pisa, I-56010 Pisa, Italy, \break
\iRAL
  Rutherford Appleton Laboratory, Chilton, Didcot, Oxon OX11 0QX United Kingdom, \break
\iRUTG
  Rutgers University, Piscataway, New Jersey 08855, \break
\iSLAC
  Stanford Linear Accelerator Center, Stanford University, Stanford, California 94309, \break
\iSOGA
  Sogang University, Seoul, Korea, \break
\iSOONG
  Soongsil University, Seoul, Korea 156-743, \break
\iTENN
  University of Tennessee, Knoxville, Tennessee 37996, \break
\iTOHO
  Tohoku University, Sendai 980, Japan, \break
\iUCSB
  University of California at Santa Barbara, Santa Barbara, California 93106, \break
\iUCSC
  University of California at Santa Cruz, Santa Cruz, California 95064, \break
\iUVIC
  University of Victoria, Victoria, B.C., Canada, V8W 3P6, \break
\iVAND
  Vanderbilt University, Nashville,Tennessee 37235, \break
\iWASH
  University of Washington, Seattle, Washington 98105, \break
\iWISC
  University of Wisconsin, Madison,Wisconsin 53706, \break
\iYALE
  Yale University, New Haven, Connecticut 06511. \break

\rm
%

\end{center}

\hfill
\end{document}